\documentclass[aps,prl,twocolumn,showpacs,amsmath,amssymb,nofootinbib,eqsecnum, preprintnumbers]{revtex4-1} 

\usepackage{color}

\newcommand{\eps}{\varepsilon}

\newcommand{\be}{\begin{equation}}
\newcommand{\ee}{\end{equation}}
\newcommand{\ba}{\begin{eqnarray}}
\newcommand{\ea}{\end{eqnarray}}
\newcommand{\nn}{\nonumber}

\def\ben{\begin{equation}}
\def\een{\end{equation}}
\def\half{\frac{1}{2}}
\def\bea{\begin{eqnarray}}
\def\eea{\end{eqnarray}}

\def\nn{\nonumber}
\begin{document}
\newcount\hour \newcount\minute
\hour=\time  \divide \hour by 60
\minute=\time
\loop \ifnum \minute > 59 \advance \minute by -60 \repeat
\def\nowtwelve{\ifnum \hour<13 \number\hour:
                      \ifnum \minute<10 0\fi
                      \number\minute
                      \ifnum \hour<12 \ A.M.\else \ P.M.\fi
	 \else \advance \hour by -12 \number\hour:
                      \ifnum \minute<10 0\fi
                      \number\minute \ P.M.\fi}
\def\nowtwentyfour{\ifnum \hour<10 0\fi
		\number\hour:
         	\ifnum \minute<10 0\fi
         	\number\minute}
\def \now {\nowtwelve}

\begin{flushright}
\hfill UPR-1267-T
\end{flushright}

\title{Thermodynamics of Asymptotically Conical Geometries}

\author{Mirjam Cveti\v c}
 \affiliation{Department of Physics and Astronomy,
 University of Pennsylvania, Philadelphia, PA 19104, USA 
\\ \& Center for Applied Mathematics and Theoretical Physics,
University of Maribor, Maribor, Slovenia}
\author{Gary W. Gibbons}
\affiliation{DAMTP, University of Cambridge, Wilberforce Road, Cambridge CB3 0WA, UK\\ 
\& Department of Physics and Astronomy,
 University of Pennsylvania, Philadelphia, PA 19104, USA\\
\& LE STUDIUM, Loire Valley Institute for Advanced Studies,
Tours and Orleans, France\\
\& 3Laboratoire de Math´ematiques et de Physique
Th´eorique, Universit´e de Tours, France}
\author{Zain H. Saleem}
\affiliation{Department of Physics and Astronomy,
 University of Pennsylvania, Philadelphia, PA 19104, USA\\
 \& National Center for Physics, Quaid-i-Azam university, Shahdara valley road, Islamabad, Pakistan}

\date{\today}

\begin{abstract}
We study the thermodynamical properties of a class of asymptotically conical (AC) geometries known as "subtracted geometries". We derive the mass and angular momentum from the regulated Komar Integral and the Hawking-Horowitz prescription and show that they are equivalent. By deriving the asymptotic charges  we show that the Smarr formula and the first law of thermodynamics hold. We also propose an analog of Christodulou-Ruffini inequality. The analysis  can be generalized to other AC geometries.
\end{abstract}

\preprint{}

\maketitle

\section{Introduction}
Black holes behave as thermodynamic objects. The thermodynamic properties of black holes are determined by the behavior of their geometry at the asymptotics due to the nature of spacetime curvature. The case of black holes in asymptotically flat spacetimes is very well understood \cite{Hawking:1973uf} and is straightforward. On the other hand, the case of black holes in asymptotically  non-zero  negative cosmological constant  (anti-deSitter (AdS)  spacetime)   possess new thermodynamic features, crucial in studies of 
gravity/field theory duality.  In general,  in fundamental theories where physical constants such as Yukawa couplings, gauge coupling constants or Newton constant  as well as  the cosmological constant  arise as vacuum expectation values of scalar fields and hence can vary, the thermodynamic laws are changed to include these variations  (see, e.g., \cite{Cvetic:2010jb}). 
These "constants" are thought of as the  vacuum expectation values of fields at asymptotic infinity, and their variation can lead to new insights into thermodynamic behavior of gravitational systems, which can play an important role in the study of gravity/field theory duality.

In this letter we focus on the study of thermodynamic properties of geometries which are asymptotically conical  (AC). The fields supporting such geometries, instead of becoming constant at spatial infinity, vary as a function of radial distance at infinity. These geometries  have very different asymptotic structure compared to the asymptotically flat and asymptotically AdS case. Their thermodynamics has not been explored  in detail,   and new insights there would provide a starting point for the study of gravity/field theory duality  for AC spacetimes.  The spacetime metrics  have the asymptotic form:
\ben
ds ^2 = - Y^{2p} dt^2+B^2 dY^2 + Y^2 \bigl(d \theta ^2 + \sin ^2 \theta d \phi ^2  \bigr )\, ,  \label{conical1}     
\een
where $p$ and $B$ are constants.
These AC metrics have the special properties that their radial distance $BY$ is a non-trivial multiple of the area distance $Y$ and their spatial metric restricted to the equatorial plane is that of a  flat two-dimensional cone. The energy density of such metrics typically falls off as inverse squared of the radial distance and thus the geometry cannot have a finite total energy. Bisnovatyi-Kogan-Zeldovich's gas sphere \cite{B-KZ,B-KT}, Barriola-Vilenkin Global Monopole \cite{Barriola}, the near horizon geometry of an extreme black hole in Einstein-Dilaton-Maxwell gravity, a black hole containing a global monopole and the cosmic string metric outside the string are all examples of asymptotically conical metrics. In this letter we  study the thermodynamics of a special  class  of metrics of the asymptotic form \eqref{conical1}, known as the "subtracted geometries" with $p=3$, $B=4$, and $Y= (8m^3r ) ^{\frac{1}{4} }$. The thermodynamics of these geometries is not known and we  show that the subtleties lie in deriving the mass, asymptotic charges and gauge fields there. The conclusions here are generalizable to other cases of AC geometries and thus are of broader interest.

Subtracted geometries are asymptotically conical black hole  solutions of N=2 supergravity theories coupled to three vector super multiplets \cite{Cvetic:2011hp,Cvetic:2011dn}. (For further details  see \cite{Cvetic:2014ina} and references therein.) They are obtained when one omits certain terms from the warp factor of the metric of the general, asymptotically flat  black hole solutions of the same theory. The radial part of the separable wave equation of a minimally coupled scalar field can now be solved by hypergeometric functions and thus possesses conformal symmetry. Furthermore, the entropy, temperature and angular velocity of the original black hole remain unchanged. However, the subtracted geometry is asymptotically conical with a Lifshitz type symmetry (a diffeomorphism under which the pull-back metric goes into a constant  multiple of itself),  with  time and radial distance scaling differently.\footnote{ Geometries with such a non-standard scaling of time have recently attracted a lot of attention for the application of gravity/field theory duality to condensed matter systems.(See, e.g., \cite{KachruYH,DanielssonGI,BertoldiVN} and references therein.).} 
The subtracted geometries can be treated physically as black holes confined in an asymptotically conical box \cite{Cvetic:2012tr}. 
The origin of this confinement property is that the square root of the time component of the metric is directly propotional to $Y^p$. This property is similar to the confining behavior of asymptotically AdS spacetimes.

The Lagrangian density of this N=2 supergravity coupled to the three vector multiplets, also known as the STU-model is given by \cite{CCLPII}:
\bea
{\cal L}_4 &=& R\, {*{\bf 1}} - \frac{1}{2} {*d\varphi_i}\wedge d\varphi_i 
   - \frac{1}{2} e^{2\varphi_i}\, {*d\chi_i}\wedge d\chi_i \nn \\
   &-& \frac{1}{2} e^{-\varphi_1}\,
( e^{\varphi_2-\varphi_3}\, {*  F_{1}}\wedge   F_{ 1}+ e^{\varphi_2+\varphi_3}\, {*   F_{ 2}}\wedge   F_{ 2} \nn \\
 &+& e^{-\varphi_2 + \varphi_3}\, {*  {\cal F}_1 }\wedge   {\cal F}_1 + 
     e^{-\varphi_2 -\varphi_3}\, {* {\cal F}_2}\wedge   {\cal F}_2)\nn\\
     &-& \chi_1\, (  F_{1}\wedge  {\cal F}_1 + 
                   F_{ 2}\wedge  {\cal F}_2)\,,
\eea
where the index $i$ ranges over $1\le i \le 3$. The four field strengths in terms of potentials are given by: 
\bea
  F_{ 1} &= &d   A_{1} - \chi_2\, d {\cal A}_2\,,  \    {\cal F}_1 = d  {\cal A}_1 + \chi_3\, d  {\cal A}_2\,, \   {\cal F}_2 = d  {\cal A}_2\, ,
     \nn\\
  F_{ 2} &=& d  A_{ 2} + \chi_2\, d {\cal A}_1 - 
    \chi_3\, d   A_{ 1} +
      \chi_2\, \chi_3\, d  {\cal A}_2\,.
\eea
The four-charge rotating black hole metric is  \cite{Cvetic:1996kv,CCLPII} \footnote{The  four gauge potentials and three  axio-dilaton fields are given in  \cite{CCLPII}. For the subtracted geometry analysis
we can take the gauge potentials $A_1=A_2=A_3\equiv A $ for gauge field strengths    $* F_1= F_2= * {\cal F} _1\equiv F\, $  and  $A_4\equiv {\cal A}$ for  $ {\cal F}_2\equiv {\cal F}$. The gauge potential definitions of this letter differ from the ones used in \cite{Cvetic:2012tr,CCLPII} by a factor of 1/2 to comply with standard literature convention. } :
\ben
d  s^2  = -  \Delta^{-\frac{1}{2}}_0  G  
( d{ t}+{ {\cal  A}})^2 + { \Delta}^{\frac{1}{2}}_0 
(\frac{d r^2} { X} + 
d\theta^2 + \frac{ X}{  G} \sin^2\theta d\phi^2 ),\label{metricg4d}
\een
where \footnote{$s_\theta\equiv \sin\theta$, $c_\theta \equiv \cos\theta$, $s_I\equiv \sinh \delta_I$, $c_I\equiv\cosh \delta_I$, $s_{2I}\equiv \sinh 2\delta_I$, $c_{2I}\equiv\cosh 2\delta_I$, ${ \Pi}_c \equiv \prod_{I=1}^4c_I$
and ${ \Pi}_s \equiv  \prod_{I=1}^4 s_I$\, .}:
\bea
{ X} & =& { r}^2 - 2{ m}{ r} + { a}^2\, ,\cr
{ G} & = &{ r}^2 - 2{ m}{ r} + { a}^2c_\theta^2
 \, , \cr
{ {\cal A}} & =& {2{ m} { a}s_\theta^2
\over { G}}
\left[ ({ \Pi_c} - { \Pi_s}){  r} + 2{ m}{ \Pi}_s\right] d\phi\, ,
\eea
and
\bea
&&{ \Delta}_0= \prod_{I=1}^4 ({ r} + 2{ m}s_I^2)
+ 2 { a}^2c_\theta^2 
[{ r}^2 + { m}{ r}\sum_{I=1}^4s_I^2  \\
&&+ 4{ m}^2 ({ \Pi}_c - { \Pi}_s){ \Pi}_s  -  2{ m}^2 \small{\sum_{I<J<K}}
s_I^2s_J^2s_K^2]\
+{ a}^4 c_\theta^4
\, .   \nn
\eea
The physical parameters (mass $M$, angular momentum $J$ and charges $Q_I$) of the general four-charge black hole are parametrized in terms of auxiliary constants $m, a, \delta_I$ as:
\begin{equation}
 M  = {\frac{1}{4}}m\sum_{I=0}^3c_{2I} ~, Q_I  = {\frac{1}{4}}m\, s_{2I}~, J  = m\, a \,(\Pi_c - \Pi_s)~,
\end{equation}
The subtraction procedure corresponds to replacing the ``warp factor''  $\Delta_0$ with $\Delta$, where:
\ben 
\Delta 
= (2m)^3 r (\Pi^2_c - \Pi^2_s) + (2m)^4 \Pi^2_s - (2ma)^2 (\Pi_c-\Pi_s)^2 c_\theta^2
\, ,
\label{deltag}\een
while keeping everything else unchanged. Importantly, this leaves the global structure unchanged, with  two horizons at $r_\pm=m\pm\sqrt{m^2-a^2}$, with the same area and surface gravity  there. (For the most general rotating black holes  of the STU model with one more independent charge parameter \cite{Chow:2013}, the subtracted geometry was obtained and analyzed in \cite{Cvetic:2014sxa}.)Therefore the entropy $S$, temperature $T$ and angular potential $\Omega$ of the subtracted geometry remain the same as in the full geometry and are given by:
\bea
S  =&& {2\pi m} \left[ (\Pi_c + \Pi_s)m + (\Pi_c - \Pi_s)\sqrt{m^2-a^2} \right]\, ,\nn\\
\quad T= &&\frac{\kappa_+}{2 \pi}\, ,\ 
\Omega =\kappa_{+} {a \over\sqrt{m^2 - a^2}}\, ,\eea
where 
\be
\frac{1}{\kappa_{+}}=2m\left[ \frac {m}{\sqrt{m^2-a^2}}(\Pi_c + \Pi_s) + (\Pi_c - \Pi_s)\right]\, .\label{kappa}
\ee
The values of the fields sourcing the subtracted geometry are however changed. The gauge fields at $\theta =0$ at the outer horizon $r_+$ are given by: \bea
{\cal A}{(r_+)}=&& \frac{2 m^2 [(2m)^2 \Pi_c\Pi_s + a^2 (\Pi_c-\Pi_s)^2]}{ R^4}\, dt\, , \\
A{(r_+)}=&& \frac{m-r_+}{4m(\Pi_c^2-\Pi_s^2)^{\frac{1}{3}}}\,dt \cr
  -&&\frac{m \, a^2\, (\Pi_c-\Pi_s)[\,r_+\,(\Pi_c-\Pi_s)+2m\Pi_s\,]\,}{ (\Pi_c^2-\Pi_s^2)^{\frac{1}{3}} R^4}\, dt\, , \nn \label{potentials}
\eea
where at the outer horizon $r_+=m+\sqrt{m^2-a^2}$,  $R^4=(2m)^2[m(\Pi_c+\Pi_s)+\sqrt{m^2-a^2}(\Pi_c-\Pi_s)]^2$. The gauge of these gauge fields is uniquely fixed by the scaling limit discussed below. (The gauge potentials  at  $\theta\ne 0$ can be found in \cite{Cvetic:2012tr}.) The three dilatons and axions are given by,
\be
e^{\varphi}\equiv e^{\varphi_{1,2,3}}  = \frac{(2m)^2(\Pi_c^2-\Pi_s^2)^{\frac{2}{3}}} {R^2}\, ,  \label{dilatons}
\ee
and 
\be
\chi\equiv {\chi_{1}} =-{\chi_{2}}={\chi_{3}} =- \frac{a {( \Pi_c-\Pi_s)}^{1/3} c_{\theta}} {2 m}\, ,  \label{dilatons}
\ee
respectively. These three axio-dilaton  fields are also fixed by the scaling limit. The asymptotic charges can then be easily obtained from the Gauss law,
\begin{equation}
{\cal{Q}} = \frac{m \Pi_c \Pi_s }{\Pi_c^2-\Pi_s^2}\, , \qquad Q= m (\Pi_c^2-\Pi_s^2)^{\frac{1}{3}} \, .
\label{charges}
\end{equation}
An important point to notice here is that the dilatons have a spatial dependence. This forces the gauge coupling constants to run logarithmically in the radial direction not even stabilizing at infinity. This is an important feature that the subtracted geometries share with Dilaton-Maxwell theory when a limit of vanishing Newton’s constant is taken.
\section{Thermodynamics}
The definitions of mass and angular momentum are heavily dependent on the asymptotics of the curved geometry. Let us start by studying the mass of our subtracted geometry first. We can afford here to deal with the static case $a=0$ since mass is defined independent of rotation. We can parameterize our static geometry by:
\begin{equation}
ds^2= -N^2 dt^2 + N^{-2} dr^2 + R^2 ( d\theta^2 + \sin^2 \theta d \phi^2)\, ,
\end{equation}
where $N={X^{\frac{1}{2}}}{\Delta^{-\frac{1}{4}}}$ and $R=\Delta^{\frac{1}{4}}$. In the Hawking-Horowitz prescription \cite{Hawking:1995fd} the mass is given by:
\begin{equation}
M_{HH}= -\frac{1}{8 \pi} \int_{S_t\to {\infty}} \, N ( K- K_0)\,  d\Omega\, ,
\end{equation}
where $d\Omega= R^2 \sin \theta d\theta d\phi$, $K$ is the extrinsic curvature of the boundary two sphere and $K_0$  in our case, will be the extrinsic curvature of the 
two-sphere \emph{embedded in asymptotically conical geometry}. Up to $O(r^{-1})$ corrections we can show that $N \sim{r}{R_0^{-1}}$, $S_t \sim 4 \pi R_0^2$,
$ R\sim R_0( 1+ {m\Pi_s^2 }([2r(\Pi_c^2 -\Pi_s^2)]^{-1})\, $ and $R_0\equiv (2m)^{\frac{3}{4}} r^{\frac{1}{4}} (\Pi_c^2-\Pi_s^2)^{\frac{1}{4}}$.
Calculating the Hawking-Horowitz mass, we get:
\begin{equation}
M_{HH}= \frac{m}{4}  \frac{\Pi_c^2+\Pi_s^2}{\Pi_c^2-\Pi_s^2}\, .
\end{equation}
Next we would like to check that the Komar mass formula gives us the same results as the Hawking-Horowitz formalism. The Komar mass is defined as:
\begin{equation}
M_K= -\frac{1}{8\pi} \int_{{S_t\to \infty}} \star d \zeta_{(t)}\, ,
\end{equation}
where $\zeta_{(t)}$ is the time-like Killing vector. In the static subtracted geometry $\zeta_{(t)}$ is given by $-{X}{\Delta^{-\frac{1}{2}}} dt$. One  can show that:
\begin{equation}
M_K= \frac{3}{4}r  - \half \frac{m \bigl( \Pi_c^2-2 \Pi_s^2 \bigr )}{\Pi_c^2-\Pi_s^2}+ O(\frac{1}{r^2})\, , \end{equation}
which  diverges linearly with $r$. The appearance of this divergence is one of the most important features that separates the asymptotically conical case with the asymptotically flat case. We can however show that this divergence gets regulated once we take the asymptotic  gauge fields and charges into account. Defining, $H^{\mu \nu}= \nabla ^\mu \zeta_{(t)}^\nu- \nabla ^\nu \zeta_{(t)}^\mu$ allows us to show that:
\be
\nabla_\mu H^{\mu \nu} = 
- 16 \pi (T^\nu_\mu - \textstyle{\frac{1}{2}} T\delta ^\nu _\mu) \zeta_{(t)}^\mu \, .
\ee
Using the above relation for the case of static electrically charged subtracted geometry, we obtain:
\begin{equation}
\nabla _ r \left( H^{r t}   + 8\pi [3 e^{\varphi} F^{r t} { {A}} _t{(r)}   +   e^{-3\varphi} {\cal F}^{r t} {\cal {A}} _t{(r)}] \right) =0\, ,
\end{equation}
As $S_t\to \infty$, ${\cal {A}} _t{(r)}\to 0$ and thus only  the term with ${{A}} _t{(r)}$ contributes. Furthermore we can identify that $ R^2 e^{\varphi}F^{rt} = { Q}$, and thus obtain:
\begin{equation}
M_{K_{reg}}= M_K + 3Q{{A}} _t{(r)}= M_{K}(r) + \textstyle{\frac{3}{4}}(m-r)\, , 
\end{equation}
where $M_K$ is the unregulated Komar mass. Therefore the terms linear in $r$ cancel and the regulated Komar mass is:
\begin{equation}
 M_{K_{reg}}= M_{HH} \, ,
\end{equation}
i.e. the Komar formula gives the same result as the one obtained through the Hawking-Horowitz  formalism. 

We can also write the explicit expression of $M_{HH}$ in terms of the reducible mass $M_{\rm{irr}}^2\equiv\frac{S}{4\pi}$,  $\cal Q$ and $Q$:
\be
M_{HH}=\frac{1}{4}\left(\frac{M_{\rm{irr}}^4}{Q^3}+\frac{{\cal Q}^2Q^3}{M_{\rm{irr}}^4}\right)\, . \label{massi}
\ee
This formula can be used to give an analogue of the Christodoulou-Ruffini inequality (\ref{massi1}) for our case, telling us the bound on how much mass of the black hole can be converted into energy.

Now we can move to define the angular momentum and study the $a\neq0$ case. In the Hawking-Horowitz formalism the angular momentum is given by:
\begin{equation}
J_{HH}= -\frac{1}{8\pi}  \int_{S_t\to\infty} ( K_{ab} - K h_{ab}) N^a {r}^b d\Omega
\, .\label{HHJ}
\end{equation}
where $a,b$ run over $ r, \theta, \phi$. $h_{ab}$ is the induced metric on the constant time hypersurface, $N^a$ is the shift vector and ${ r}^a$ is the unit normal to the boundary two sphere. In the axially symmetric case of subtracted geometry the second term does not contribute because $h_{ab}$ does not have a $\phi r$-component. On the other hand the Komar integral for the angular momentum is given by: 
\begin{equation}
J_K= \frac{1}{16\pi} \int_{S_t\to{\infty}} \nabla^{\mu} \zeta^{\nu}_{(\phi)}d S_{\mu \nu}\, , \label{KJ}
\end{equation}
where $\zeta^{\mu}_{(\phi)}$ is the rotational Killing vector and the area element $dS_{\mu\nu}= -2 n_{[\mu} r_{\nu]} d\Omega$ with $ n^{\mu}=e^{\mu}_a n^a$ and $ r^{\mu}=e^{\mu}_a r^a $ being the time-like and space-like normals to the surface $ S_t$.
We can show the equality of  (\ref{HHJ}) and (\ref{KJ}) by employing:
\begin{eqnarray}
\nabla^{\mu} \zeta^{\nu}_{(\phi)} n_{\mu} r_{\nu}&= & K_{ab} N^a n^b\, .
\end{eqnarray}
It is a simple exercise to show:
\begin{equation}
J_{HH}=J_K=J\equiv am ( \Pi_c-\Pi_s)\, .
\end{equation}
Once we have defined mass and angular momentum we can easliy show that the Smarr law:
\begin{equation}
 M_{HH}=2TS +{{\cal {A}} _t{(r_+)}}{\cal{Q}}+3{{{A}} _t{(r_+)}}Q + 2J \Omega\label{smarr}\, ,
\end{equation}
and the first law of thermodynamics:
\begin{equation}
dM_{HH}= T\, dS +{{\cal {A}}_t{(r_+)}}\, d{\cal{Q}}+3{{A}}_t{(r_+)}\, dQ + \Omega\, dJ\, ,  \label{first}
\end{equation}
continue to hold for our geometry. This gives us further confidence in our definitions of the mass and angular momentum. Another important point to notice is that in order for these laws to be obeyed, the gauge fixing of the fields was crucial and was uniquely fixed by the scaling limit.

\section{Scaling Limit} 
There are two ways to obtain the subtracted geometries starting from the original ones. Firstly they can be obtained by applying the Harrison transformations \cite{Cvetic:2012tr,Virmani:2012kw,Cvetic:2013cja} and secondly it can also be obtained via the scaling limit \cite{Cvetic:2012tr}. In this section we apply the scaling limit to the mass and angular momentum formulae obtained in the original black hole calculations and see how they agree with the subtracted geometry answers that we obtained above by direct calculations. The limit is implemented by means of the following scalings:
\begin{eqnarray}
&& m\to m\, \epsilon,\,\,\,\, \ r \to r\, \epsilon,\,\,\,\,     t\to  t\,\epsilon^{-1},\,\,\,\, a \to a\, \epsilon\,, \\ 
&& \sinh^2\delta_{1,2,3}\to \frac{ (\Pi_c^2-\Pi_s^2)^{\frac{1}{3}}}{ \epsilon^{\frac{4}{3}}}, \sinh^2\delta_4\to \frac {\Pi_s^2}{(\Pi_c^2-\Pi_s^2)}\,. \nn
\label{scaling}
\end{eqnarray}
The scaling limit ensures that the entropy, the surface gravity, the angular velocity {\it and} the angular momentum are the same as those of the asymptotically flat black hole, with the  result for the angular momentum confirmed  by the independent calculation above. 
The matching of the mass formula is however more involved. The mass of the original four-charge STU black hole is given by:
$M= \textstyle{\frac{1}{4}} m \sum_I \cosh 2 \delta_I\, .
$
Its scaling limit is:
\begin{equation}
\frac{ M }{ \eps}=  \frac{3m }{2} \frac{ (\Pi_c^2- \Pi_s^2)^{\frac{1}{3}}}{\eps^{\frac{4}{3}}}+\frac{m}{4} \frac{ \Pi_c^2 + \Pi_s^2}{ \Pi_c^2-\Pi_s^2}+ \frac{3m}{4}\, .  \label{scalingmass}
\end{equation}
In the scaling limit the gauge potentials $A_{1,2,3}$ acquire an infinite constant term,  not included in (\ref{potentials}),  which along with the large charges $Q_{1,2,3}$ ensure that in the Smarr relation, the product of original charges $Q_i=m\sinh2\delta_i$ with $A_i$ ($i=1,2,3$) contain  divergent parts which cancel the divergent part in (\ref{scalingmass}).  Furthermore the constant term $\frac{3m}{4}$ in (\ref{scalingmass}) is cancelled by a product of the sub-leading contribution in $Q_{1,2,3}$  and the divergent part of $A_{1,2,3}$. The remaining contributions are due to the precisely  quoted charges  (\ref{charges}) and gauge potentials (\ref{potentials}), thus verifying that the mass of the subtracted geometry is indeed $M_{HH}$.
\section{Conclusion and Discussion}
It should be noted that 
our successful
extension of a coherent black hole thermodynamic theory involving 
appropriately
defined asymptotic charges to the case of
subtracted geometries
depends crucially on taking seriously their asymptotically conical 
nature. This differs both qualitatively and quantitatively from the
the standard asymptotically flat and asymptotically AdS cases. 
Nevertheless the
end result shares the universal features of those cases and gives 
further
support to the idea that there are microscopic states or degrees of 
freedom (possibly stringy)
counted by the entropy of black holes and the number of such states
is insensitive to which environment they find themselves in.

Our analysis also resulted in the explicit expression  (\ref{massi})  for  $M_{HH}$ in terms of $M_{\rm{irr}}$, $\cal Q$ and $Q$. This expression lends itself to propose an analog of the Christodulou-Ruffini inequality \cite{Christodoulou:1972kt}:
\be
M_{HH}\geq \frac{1}{4}\left(\frac{M_{\rm{irr}}^4}{Q^3}+\frac{{\cal Q}^2Q^3}{M_{\rm{irr}}^4}\right)\, . \label{massi1}
\ee
Such an inequality can be tested, at least in time symmetric data context  \cite{Cvetic:2014vsa} by taking the scaling limit of the initial data results for the STU model. 

Furthermore for these initial data, the Einstein-Rosen Bridge structure is manifest from eq.(2.24) and eq.(2.25)
of \cite{Cvetic:2014ina} where 
the reflection map  of Kruskal-Szekeres coordinates $(U,V) \rightarrow (-U,-V)$
is an isometry  that leaves the radial coordinate $r$ invariant but fixes the $U=const.V$
surfaces, which in regions I and IV are constant time surfaces.
Thus the initial data of the asymptotically conical 3-metrics has to be joined by an Einstein-Rosen throat. 
Further study of these properties of the subtracted geometry is of great interest.\\
\vskip 0.1 in
\noindent{\bf Acknowledgements} 
We would like to thank Finn Larsen, Christopher Pope, Maria Rodriguez, Alejandro Satz and Oscar Varela for discussions. The work is supported in part by the DOE Grant DOE-EY-76-02- 3071 (MC), the Fay R. and Eugene
L. Langberg Endowed Chair (MC),  Simons Foundation Fellowship (MC) and the Slovenian Research Agency (ARRS) (MC).


\providecommand{\href}[2]{#2}

\begingroup\raggedright

\end{document}